\documentclass[twocolumn,aps,prl,floatfix,10pt,groupedaddress,longbibliography]{revtex4-1}
{}
\usepackage{xcolor}
\usepackage[utf8]{inputenc} 
\usepackage{bm}
\usepackage{amsmath,amsthm,amssymb}
\usepackage{graphicx} 
\usepackage{tikz,tkz-tab}{{}}
\usepackage[
  colorlinks=true,
  urlcolor=blue,
  linkcolor=blue,
  citecolor=blue
]{hyperref}
\usepackage[normalem]{ulem}

\def\url#1{}

\newcommand{\tmel}{t_\mathrm{mol}}

\long\def\soutjd#1{}

\begin{document}

\title{Probing spin correlations in a Bose--Einstein condensate near the single atom level}

\author{An Qu, Bertrand Evrard, Jean Dalibard and Fabrice Gerbier}

\affiliation{Laboratoire Kastler Brossel, Coll{\`e}ge de France, CNRS, ENS-PSL Research University, Sorbonne Universit{\'e}, 11 Place Marcelin Berthelot, 75005 Paris, France}

\date{\today}
\pacs{}

\begin{abstract}
Using parametric conversion induced by a Shapiro-type resonance, we produce and characterize a two-mode squeezed vacuum state in a sodium spin 1 Bose--Einstein condensate. Spin-changing collisions generate correlated pairs of atoms in the $m=\pm 1$ Zeeman states out of a condensate with initially all atoms in $m=0$. A novel fluorescence imaging technique with sensitivity $\Delta N \sim 1.6$ atom enables us to demonstrate the role of quantum fluctuations in the initial dynamics and to characterize the full distribution of the final state. Assuming that all atoms share the same spatial wave function, we infer a squeezing parameter of 15.3\,dB.
\end{abstract}

\maketitle


\paragraph{Introduction.}
Entanglement between subsystems is both an essential concept for the understanding of quantum physics and a unique resource for emerging quantum technologies \cite{Leggett2002,haroche2006,arndt2014}. For example in metrology, one can exploit quantum correlations between particles to improve interferometric measurements\,\cite{Giovannetti2004a,pezze2018a}. 
Instead of the standard quantum limit where the sensitivity scales as $1/\sqrt{N}$ for an ensemble of $N$ uncorrelated particles, interferometry with entangled states can in principle reach the Heisenberg limit scaling  as $1/N$, a potentially very large gain. 

Among the several kinds of entangled states that can be used for quantum metrology\,\cite{pezze2018a}, the two-mode squeezed vacuum (TMSV) state is particularly interesting. It corresponds to a superposition of twin Fock states with exactly the same number of particles in modes $a,b$. A measurement of the occupation number $N_a$ for the mode $a$ determines exactly $N_b$ for the mode $b$, allowing, for example, the detection of absorption processes at the single-particle level. TMSV states have been produced in several  platforms: spontaneous parametric down-conversion in quantum optics\,\cite{walls}, superconducting circuits\,\cite{nation2012colloquium}, coherent collisions in a Bose--Einstein condensate (BEC)\,\cite{law1998a,duan2000b,pu2000a,sorensen2001a,duan2002a,mias2008a,leslie2009a,
klempt2010a,gross2011a,bookjans2011a,lucke2011a,hamley2012a,Luecke2014a,hoang2016a,luo2017a,fadel2018a,kunkel2018a,lange2018a}.  Early studies have explored the potential of TMSV states for interferometry, finding them suitable to reach the Heisenberg limit\,\cite{holland1993a,bouyer1997a,kim1998a,Dunningham2002a,pezze2018a}. Beyond metrology, TMSV states are essential for photonic quantum information processing\,\cite{flamini2018photonic}, and may be also useful for gravitational wave detection\,\cite{aasi2013enhanced}. 

To fully characterize such states and harness their entanglement, the detection of the mode populations with single-quantum resolution is paramount. For a large number of particles, this has been a long-standing obstacle both in optics and atomic physics. For atomic systems, the detection noise reported for entangled state production ranged from several particles\,\cite{muessel2013a} to several tens\,\cite{gross2011a,bookjans2011a,lucke2011a,hamley2012a,Luecke2014a,hoang2016a,luo2017a}. Single-atom sensitivity was demonstrated for a $\sim 10^3$ atom cloud recaptured in a magneto-optical trap \cite{hume2013a} but only for the total population. Resolving the individual mode populations does not seem reachable with this technique. In this Letter, we take advantage of the recently demonstrated atomic Shapiro resonance\,\cite{evrard2019a} to generate a TMSV state in a spinor BEC of sodium atoms (spin 1). Modes $a,b$ correspond to the magnetic sublevels $m=\pm 1$, which allows us to use a Stern-Gerlach splitting followed by a high-precision fluorescence imaging for atom counting, with a sensitivity of about $1.6$ atom per spin component. Assuming that all atoms occupy the same spatial mode, we demonstrate a detection-limited compression of $15.3$\,dB. 

\paragraph{Parametric conversion.}
Our experiment is well described within the single-mode approximation in which all atoms share the same spatial wave function, but can form highly entangled spin states. The Shapiro resonance used in this work is essentially equivalent to the well-known parametric conversion process in optics\,\cite{walls,nation2012colloquium}. The initial state consists in having all atoms in $m=0$ and can be viewed as a ``vacuum state". The parametric conversion generates entangled pairs of $m=\pm 1$ atoms by the coherent spin-changing collisional process $2 \times (m=0) \to (m=+1)+(m=-1)$\,\cite{duan2000b,pu2000a,sorensen2001a,klempt2010a,gross2011a,bookjans2011a,lucke2011a,hamley2012a,Luecke2014a,hoang2016a,luo2017a}. 

The main physical process can be explained by treating the highly populated $m=0$ mode as a classical source\,\cite{duan2000b,pu2000a,sorensen2001a}. The  Hamiltonian modeling the parametric conversion process is 
\begin{align}\label{eq:Hparam}
\hat{H}_\mathrm{prm}&=K \left(\hat{a}_{+1}^\dagger \hat{a}_{-1}^\dagger+\mathrm{H.c.}\right),
\end{align}
with $\hat{a}_m^\dagger$ the creation operator for component $m$. The initial vacuum state $\vert 0, 0 \rangle$ evolves into the TMSV state
\begin{align}\label{eq:TMSS}
\vert \Psi(t) \rangle= {\sqrt{1-\vert \eta \vert^2}}\;\sum_{k=0}^{N/2} \eta^k \vert k, k \rangle,
\end{align}
where $\vert k,k \rangle$ denotes the Fock state with $N_{+1}=N_{-1}=k$, $N_0=N-2k \approx N$, and   $\eta(t)=-{\rm i}\tanh (Kt/\hbar)$. The properties of the TMSV state (\ref{eq:TMSS}) are best discussed by introducing the magnetization and pair number operators
\begin{align}\label{eq:JzNp}
\hat{J}_z &= \frac{1}{2} \left(\hat{N}_{+1} - \hat{N}_{-1}\right),\quad 
\hat{N}_\mathrm{p} = \frac{1}{2} \left(\hat{N}_{+1} + \hat{N}_{-1}\right),
\end{align}
with $\hat N_m=\hat a_m^\dagger \hat a_m$.
The number of pairs obeys a Bose--Einstein distribution with the time-dependent mean $\bar{N}_\mathrm{p}= \vert \eta \vert^2/(1-\vert \eta \vert^2)$ and variance $\Delta N_{\rm p}^2=\bar{N}_\mathrm{p}(\bar{N}_\mathrm{p}+1) \sim \bar{N}_\mathrm{p}^2$. On the other hand, the magnetization $J_z$ remains exactly equal to zero at all times, corresponding to perfect squeezing. 


\paragraph{Shapiro resonance.}

\begin{figure}[t]
  \includegraphics[width=3.375in]{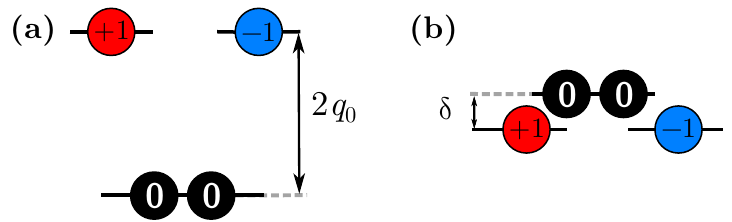}
  \caption{\textbf{Shapiro resonance for a spin 1 BEC.} 
  \textbf{(a)} The quadratic Zeeman energy originating from a static magnetic field creates an energy offset $2q_0>0$ between a $(+1,-1)$ pair and a $(0,0)$ one. 
  \textbf{(b)} In the presence of an additional modulated field, the dynamics can be described by the secular Hamiltonian (\ref{eq:Hsec}). The offset $2q_0$ is replaced by the detuning $\delta$ from the resonance, allowing one to control the sign of the energy difference between $(+1,-1)$ and  $(0,0)$ pairs. }
  \label{fig:basic_process}
\end{figure}

Our experiment is performed with a BEC of $N \approx 2700$ atoms in the single-mode regime, with all three Zeeman components of the $F=1$ hyperfine level trapped identically in a crossed optical dipole trap\,\cite{jacob2011a,jacob2012a,zibold2016a}. In order to minimize the effect of residual magnetic fluctuations, we apply a static bias field $B_0\approx0.46\,$G. It raises the Zeeman energy of a  $(+1,-1)$ pair by the amount $2q_0$ above the energy of two $m=0$ atoms, where $q_0\propto B_0^2$ is the quadratic Zeeman shift for a $m=\pm 1$ state (figure \ref{fig:basic_process}\textbf{a}). This Zeeman shift thus puts out of resonance the process described by Eq.\,(\ref{eq:Hparam}). In addition, s-wave interactions for sodium atoms in the $F=1$ level are antiferromagnetic, which increases the energy difference between $(+1,-1)$ and $(0,0)$. In order to restore the resonance for the parametric process, one could think of  differentially shifting the $m=0$ and $m=\pm 1$ states using a microwave coupling to the $F=2$ hyperfine level\,\cite{gerbier2006}. However, losses due to hyperfine relaxation collisions \cite{gorlitz2003} would constitute a significant source of decoherence for our experimental parameters. Instead we use a parametric instability resulting from a coherent drive of our spinor gas\,\cite{hoang2016a,evrard2019a}, based on a Shapiro-type resonance.  

In order to induce the Shapiro resonance, we superimpose an oscillating magnetic field $\boldsymbol B \cos(\omega t/2)$ to the static magnetic field $\boldsymbol B_0$. The directions of $\boldsymbol B_0$ and $\boldsymbol B$ are orthogonal, resulting in the quadratic Zeeman energy $q(t)=q_1+q_2 \cos(\omega t)$, with $q_1/h=268\,$Hz and $q_2/h=210\,$Hz ($q_1\propto B_0^2+B^2/2$, $q_2\propto B^2/2$). The modulation frequency $\omega/2\pi \approx 560\,$Hz is chosen close to $2q_1/h$ to induce the resonance. The response of the driven system then consists in a fast micromotion on top of a slower motion\,\cite{evrard2019a}, the latter being described by the secular Hamiltonian
\begin{align}
\nonumber
\hat{H}_\mathrm{sec}=&\hbar\delta \hat{N}_\mathrm{p} +\frac{2U_s}{N}\hat{N}_\mathrm{p}\left(N-2\hat{N}_\mathrm{p}\right)\\
\label{eq:Hsec}
&
+\frac{\kappa U_s}{N} \left(\hat{a}_{+1}^\dagger \hat{a}_{-1}^\dagger \hat{a}_0^2+\mathrm{H.c.}\right),
\end{align} 
where the operator $\hat{N}_\mathrm{p}=(N-\hat{N}_0)/2$ counts the number of $m=\pm 1$ pairs, $\hbar\delta=2q_1-\hbar\omega$ is the detuning from resonance and $U_s\approx h\times 18 \,$Hz the spin interaction energy. The secular Hamiltonian is formally similar to the Hamiltonian of a single-mode spinor BEC without modulation\,\cite{law1998a,stamperkurn2013a} with adjustable sign and strength for the quadratic Zeeman effect and for the spin-mixing interaction (figure \ref{fig:basic_process}\textbf{b}). The modulation indeed renormalizes both quantities $2q_0 \to \hbar\delta$ and $U_s \to \kappa U_s$ with $\kappa \approx 0.34$ in our experiment\,\cite{evrard2019a}.

Assuming that the $m=0$ state contains most of the population, we can simplify $\hat{H}_\mathrm{sec}$ by keeping only terms quadratic in the operators $\hat{a}_{\pm 1}$ and $\hat{a}_{\pm 1}^\dagger$, and then diagonalize $\hat{H}_\mathrm{sec}$ by a Bogoliubov transformation\,\cite{duan2000b,mias2008a}. The Bogoliubov energy is $\hbar\omega_B=\sqrt{\lambda_+ \lambda_-}$ with $\lambda_\pm=(1\pm \kappa)U_s+\hbar\delta/2$. The range of detuning
\begin{align}
-2(1+\kappa)U_s \leq \hbar\delta \leq -2(1-\kappa)U_s
\label{eq:instability}
\end{align}
corresponds to a dynamical instability window with imaginary $\omega_B$. 
Within that window, the quasiparticle operators grow exponentially at a rate $ |\omega_\mathrm{B}|$ and the evolution from the initial state $|0,0\rangle$ leads to a TMSV state exactly as for the ``ideal'' parametric amplifier described by Eq.\,(\ref{eq:Hparam}). In the following we choose $\delta/2\pi=-24\,$Hz, at the upper border of the instability window (\ref{eq:instability}).

\begin{figure}
  \includegraphics[width=3.375in]{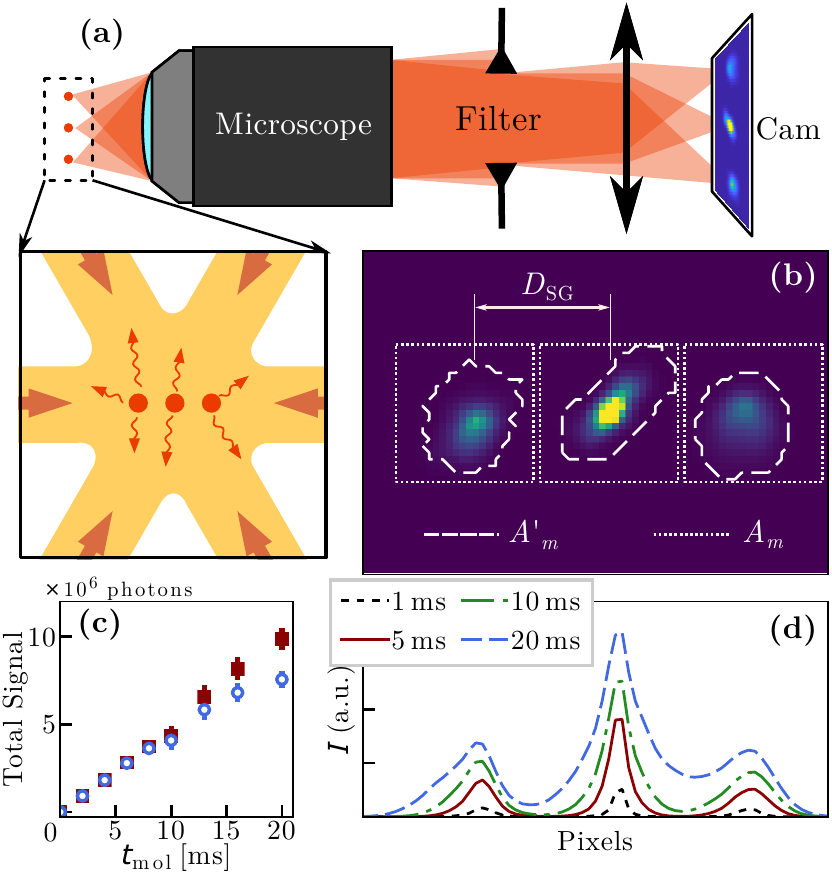}
  \caption{\textbf{Spin-resolved fluorescence imaging.} 
  \textbf{(a)} Imaging system recording the fluorescence from each Zeeman component on a CCD camera. A microscope objective (numerical aperture 0.33) is relayed by a pair of lenses (not shown) and a tailor-made spatial filter. 
  \textbf{(b)} Typical fluorescence image for a molasses duration $\tmel=5\,$ms. The distance between adjacent clouds is $1.3\,$mm. The dotted (resp.\;dashed) contours $A_m$ (resp. $A_m'$) show the raw (resp.\;optimized) regions of interest. 
  \textbf{(c)} Mean fluorescence signal for a pure $m=0$ cloud as a function of $\tmel$. The squares (resp.\;circles) indicate the total fluorescence signal in $A_0$ (resp.\;$A_0'$). 
  \textbf{(d)} Fluorescence profiles for four values of $\tmel$ after integration along the $y$ direction. }
  \label{fig1}
\end{figure}

\paragraph{Fluorescence imaging.}
In order to analyze the state produced by the parametric resonance,  we developed a  ``Stern--Gerlach fluorescence imaging'' technique to measure the populations $N_m$ of the Zeeman states (Fig.\,\ref{fig1}\textbf{a}). We first release the atoms from the trap and apply a magnetic field gradient to separate the spin components in three well-isolated clouds. Then, we switch on a three-dimensional optical molasses for a duration $\tmel$. Atoms continuously scatter photons off the red-detuned molasses beams while being simultaneously cooled. We collect part of the fluorescence light emitted by each cloud on a scientific-grade CCD camera. 

Fig.\,\ref{fig1}\textbf{b} shows a typical fluorescence image for $\tmel=5\,$ms. Using absorption imaging for global calibration, we find that $\approx 450$ photons are detected per atom. This represents $\sim 1\%$ of the total emitted fluorescence light. In order to minimize the contribution of the diffuse background light, we use as regions-of-interest (ROIs) the smallest areas $A_m'$ that contain 99\% of the total signal measured on larger areas $A_m$, as shown in Figs.\,\ref{fig1}{\bf b},{\bf c}. For each image, the mean contribution of the background light ($\sim 4.2\times 10^5$\,photons/ROI) is estimated from the signal out of the ROIs and subtracted from the total count\,\cite{SM}. 

As for the noise, the main contribution is the optical shot noise of the background light, $1.4$ times larger than the single-atom signal for $\tmel=5\,$ms. With typically 100 atoms in each state $m=\pm 1$, the shot noise of the fluorescence light is notably smaller, $0.5 \times$\,single-atom signal. Both contributions decrease in relative value for a longer exposition time $t_{\rm mol}$. A third contribution comes from atom losses during the molasses phase\,\cite{muessel2013a}, presumably because of light-assisted inelastic collisions. Losses increase with $\tmel$, leading to an optimal molasses duration that minimizes the atomic detection noise. We find the optimal choice around $\tmel=5\,$ms, leading to the noise per Zeeman component $\Delta N_m \approx 1.6\,$ atom. For this value of $t_{\rm mol}$, the overlap between the clouds is negligible as shown in Fig.\,\ref{fig1}{\bf d}: false assignment to the wrong $m$ state is less than 0.1\,\% \cite{SM}.

Single-atom detection can also be achieved with absorption imaging\,\cite{streed2012absorption}. It requires the number of photons absorbed by a single atom to overcome the shot noise in the detection of the probe beam. This condition, when applied to the large size of our clouds after Stern--Gerlach splitting, requires a large number of absorbed photons per atom. This number is reachable only if an OM cools the atoms during the imaging process itself. For our setup, the analysis of the expected signal-to-noise ratio then shows that it is  favorable to  measure the fluorescence from the (anyway necessary) OM beams.  


\paragraph{Evolution and characterization of the TMSV state.} 

\begin{figure}
  \includegraphics[width=3.375in]{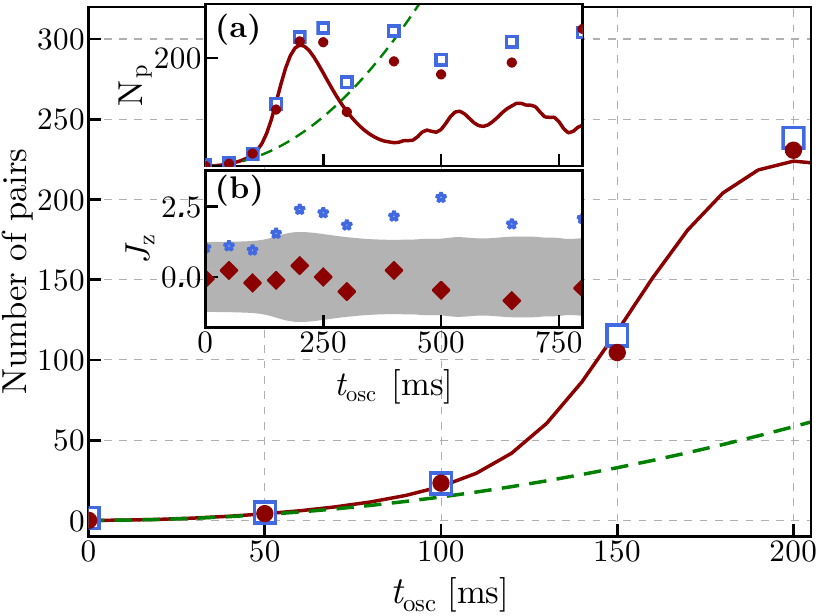}
  \caption{\textbf{Production of a TMSV state.} 
  Measured evolution of the mean number of pairs $\bar{N}_\mathrm{p}$ (red circles) and the standard deviation $\Delta N_\mathrm{p}$ (blue squares). The {continuous} red line shows the numerical solution of the Schr\"odinger equation using the secular Hamiltonian (\ref{eq:Hsec}). The {dashed} green line shows the prediction from the Bogoliubov Hamiltonian derived from $\hat{H}_\mathrm{sec}$.
  Inset {\textbf{(a):}} Same data for longer evolution times.
  Inset {\textbf{(b):}} Evolution of the average value $\bar{J}_z$(red diamonds) and standard deviation $\Delta J_z$(blue stars). }
  \label{fig3}
\end{figure}

We now describe the production of TMSV states in our setup. The initial state is obtained by a combination of evaporation and spin distillation in the presence of a magnetic gradient, and corresponds within noise to all atoms in $m=0$. More precisely an average over 1000 shots gives an initial population in the $m=\pm1$ modes compatible with zero with a standard error $\approx 0.07$ atom \cite{SM}. We can thus safely attribute the onset of the parametric instability dynamics to quantum fluctuations. 

Fig.\,\ref{fig3} shows the measured mean number of pairs and its standard deviation. At all times, the relation $ \Delta N_p(t) \approx \bar{N}_{\mathrm{p}}(t)$ expected for a Bose--Einstein distribution is well fulfilled. {\color{black} Fig.\,\ref{fig3}(\textbf{b}) further shows that the mean value of $J_z$ remains compatible with zero, and its standard deviation $\Delta J_z$ is at the level of the detection noise up to $t_{\rm osc}=150\,$ms, where $\bar{N}_{\mathrm{p}}\approx 100$. At longer times and larger $\bar{N}_{\mathrm{p}}$, we observe a small increase of $\Delta J_z$, possibly due to atom losses in the molasses. Atom losses during the preparation phase and interactions between BEC atoms and the residual thermal cloud may also play a role. In any case, our observations demonstrate the generation of correlated atom pairs as well as a strong robustness of the squeezing on a 200\,ms time scale.

We also show in Fig.\,\ref{fig3} two theoretical predictions. The first one plotted with a {continuous line} is the numerical solution of the Schr\"odinger equation with the full secular Hamiltonian $\hat{H}_\mathrm{sec}$. It reproduces remarkably well the experimental results for interaction times up to $200$\,ms, including the saturation behavior with a maximum of $\sim 440$ atoms converted into $220$ pairs $(+1,-1)$. At longer times, this numerical solution exhibits oscillations that are not observed experimentally, possibly  because of the decoherence/loss mechanisms mentioned above. 
The second prediction shown with a {dashed} line is obtained from the Bogoliubov Hamiltonian, when only terms quadratic in $\hat a_{\pm 1}$ are kept in $\hat{H}_\mathrm{sec}$. It agrees with the experimental results only for short interaction times ($t<50\,$ms) and underestimates the pair production beyond this point. This discrepancy originates from the evolution of the effective detuning $\delta$, which becomes more negative as the number of pairs $(+1,-1)$ increases, hence shifts deeper into the instability region (\ref{eq:instability}). This positive feedback on the pair production is properly taken into account in the full numerical solution based on $\hat{H}_\mathrm{sec}$, but is absent from its quadratic approximation\,\cite{SM}.

We performed a detailed characterization of the state produced after the evolution time $t=150\,$ms. Fig.\,\ref{fig4} shows as red dots the repartition of about 500 measurements in the $N_\mathrm{p}-J_z$ plane, along with the marginal distributions. For comparison, we also show as blue dots the measured distributions for a spin coherent state 
$\left( |m=+1 \rangle - |m=-1 \rangle\right)^{\otimes 2\bar{N}_\mathrm{p}^{\rm coh}}$ with a mean number of pairs $\bar{N}_\mathrm{p}^{\rm coh} \approx 76$. In Fig.\,\ref{fig4}\textbf{(c)}, we also plot the expected Bose--Einstein distribution ${\cal P}(N_{\rm p})$ of mean $\bar N_{\rm p}$. The experimentally measured probability distribution of $N_{\rm p}$ is in excellent agreement with this prediction.

\begin{figure}
  \includegraphics[width=3.375in]{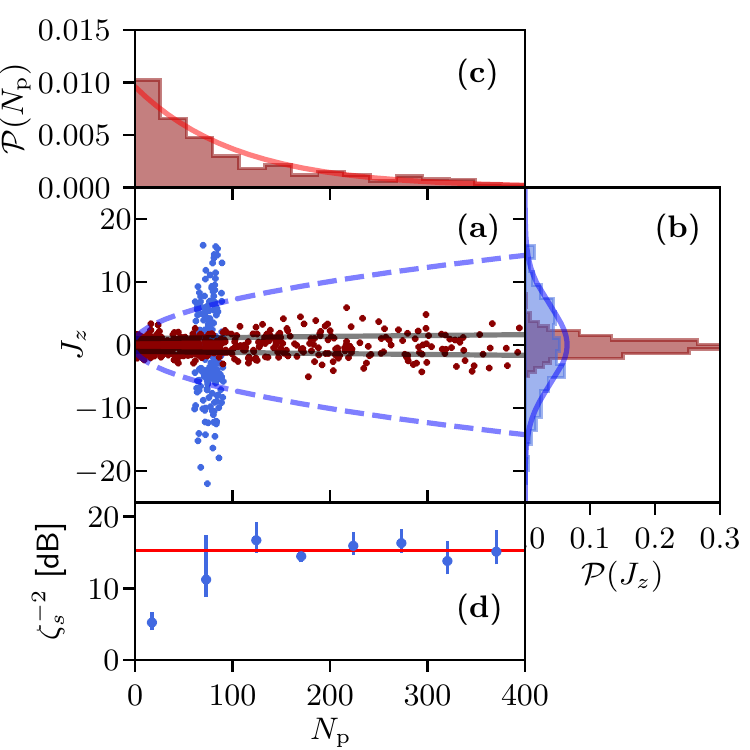}
  \caption{\textbf{Characterization of a TMSV state.}
  \textbf{(a)} The red points show 536 repeated measurements of the TMSV state produced by the sequence of Fig.\,\ref{fig3} for $t_\mathrm{osc}=150\,$ms ($\bar N_{\rm p}=105$). The blue points are experimental results for a balanced spin coherent state with $\sim 2\bar{N}_\mathrm{p}$ atoms in $m=\pm 1$. The dashed blue curve shows the variation of the typical range for $J_z$ ({i.e.}, $\pm \Delta J_z$) for a coherent state with $2N_\mathrm{p}$ atoms. The solid gray lines show the expected detection noise.  
  \textbf{(b)} Histograms of $J_z$ for the TMSV  state (red) and the coherent state (blue). The respective standard deviations are $1.55$ and $7.26$.
  \textbf{(c)} Histogram of $N_\mathrm{p}$ for the TMSV state. The solid line is the Bose--Einstein distribution of mean $\bar N_{\rm p}=105$. 
 \textbf{(d)} Spin-squeezing parameter $\zeta_s^{-2}$ versus $N_{\rm p}$, calculated with a $\Delta N_{\rm p}=50$ bin width. Here all data at $t\leq250\,$ms are used (928 measurements). The error bar (66\% confidence interval) is obtained using the bootstrap method. The average squeezing parameter (red line) for $N_{\rm p}>100$ is $\zeta_s^{-2}=15.3\,$dB.}
  \label{fig4}
\end{figure}

To characterize the entanglement of the $N_{\rm p}$ pairs of $m=\pm 1$ atoms considered as $2N_{\rm p}$ pseudo-spin $1/2$ particles, we use the spin-squeezing parameter \cite{Vitagliano:2014_PhysRevA.89.032307},
\begin{equation}
\zeta_s^2\equiv \frac{(2N_{\rm p}-1)\;(\Delta J_z)^2}{\langle J_x^2\rangle + \langle J_y^2\rangle -N_{\rm p}} \approx \frac{2 (\Delta J_z)^2}{N_\mathrm{p}}\,
\label{eq:zetas}
\end{equation}
where the second equality assumes that (i) $N_{\rm p}\gg 1$ and (ii) the pseudo-spin state is fully symmetric, which holds if all atoms share the same spatial mode \cite{SM}. Any value $\zeta_s^2<1$ signals that the pseudo-spin state is not separable. We show $\zeta_s^{-2}$ versus $N_{\rm p}$ in Fig.\,\ref{fig4}{\bf (d)}. For $N_{\rm p}$ above 100, we find  $\zeta_s^2\approx0.0293$, \emph{i.e.} a squeezing level of $15.3\,$dB.


\paragraph{Discussion and outlook.}
We have described in this Letter the production and the characterization of a TMSV state using Floquet engineering in a spinor BEC. The detection scheme uses a novel, spin-resolved fluorescence imaging technique with a sensitivity close to single-atom resolution, $\Delta N \simeq 1.6$ atoms. This sensitivity is currently mostly limited by the shot noise of the residual stray light. We are confident that it could be further improved below the single-atom level, using a dedicated shielding of the background light inside the vacuum chamber. 

Such TMSV states can be directly used for interferometric measurements at the Heisenberg limit. One can use for example a Mach--Zehnder interferometer with each mode $m=\pm 1$ injected in one of the input ports\,\cite{kim1998a}. As for a Ramsey-type experiment, two $\pi/2$ Rabi pulses between $m=\pm 1$ play the role of the entrance and exit beam splitters, and the measurement of $J_z^2$ at the output of the interferometer reveals the presence of a phase shift in one of the two arms, with an uncertainty scaling as $1/N_{\rm p}$. Here, we infer from our current detection noise a phase sensitivity of 7.6 dB beyond the standard quantum limit \cite{SM}.

Stern-Gerlach fluorescence imaging can be implemented in almost any cold atom experiment and it constitutes a convenient tool toward high-precision interferometry with spinor gases. Here we worked with a few hundred entangled particles, but the method can be generalized to larger samples, such as the $10^4$ entangled-particle sample of \cite{luo2017a}, as long as losses during the molasses phase remain small. Immediate applications of such interferometers are magnetometry and magneto-gradiometry\,\cite{vengalattore2007a,stamperkurn2013a}. Furthermore, Refs.\,\cite{fadel2018a,kunkel2018a,lange2018a} recently demonstrated that a Stern-Gerlach apparatus (or generalization thereof) was able to transfer entanglement from the spin sector to the spatial degrees of freedom. This enables a broader range of applications, including in particular inertial sensing and gravimetry.

\begin{acknowledgments}	
\noindent  We thank the members of the BEC group at LKB for insightful discussions. This work was supported by ERC (Synergy Grant UQUAM). LKB is a member of the SIRTEQ network of R\'egion Ile-de-France. A.Q. and B.E. contributed equally to this work.
\end{acknowledgments}	

\bibliographystyle{apsrev}
\bibliography{fluoBib}

\end{document}